\documentclass[a4paper,twocolumn,superscriptaddress,11pt]{quantumarticle}
\pdfoutput=1
\usepackage[utf8]{inputenc}
\usepackage[english]{babel}
\usepackage[T1]{fontenc}
\usepackage{graphicx,bbm,amsfonts,amsmath,amsthm,mathrsfs,amssymb}
\usepackage[usenames,dvipsnames]{color}
\usepackage[numbers,sort&compress]{natbib}

\newcommand{\id}{\mathbbm{1}}
\newcommand{\transpose}{{\mkern-1.5mu\mathsf{T}}}
\newcommand{\dd}{\mathrm{d}}
\newcommand{\tr}{\mathrm{tr}}
\newcommand{\Tr}{\mathrm{Tr}}
\newcommand{\vc}{\mathrm{vec}}
\newcommand{\avg}[1]{\langle{#1}\rangle}
\newcommand{\avgS}[1]{\avg{#1}_\mathrm{sq}}
\newcommand{\norm}[1]{\Vert #1\Vert}

\newcommand{\vb}[1]{\mathbf{#1}}
\newcommand{\ket}[1]{|#1\rangle}
\newcommand{\bra}[1]{\langle#1|}
\newcommand{\bbra}[1]{\langle\!\langle{#1}\vert}
\newcommand{\kket}[1]{\vert{#1}\rangle\!\rangle}
\newcommand{\bbkk}[2]{\langle\!\langle{#1}\vert{#2}\rangle\!\rangle}
\newcommand{\rbra}[1]{({#1}\vert}
\newcommand{\rket}[1]{\vert{#1})}
\newcommand{\rbrk}[2]{({#1}\vert{#2})}
\newcommand{\rdyad}[2]{\vert{#1})({#2}\vert}
\newcommand{\idDyad}{\overset{\leftrightarrow}{\mathbf{1}}}

\newcommand{\upi}{\mathrm{i}}
\newcommand{\upe}{\mathrm{e}}
\newcommand{\upu}{\mathrm{u}}

\newcommand{\cB}{\mathcal{B}}

\newcommand{\cE}{\mathcal{E}}
\newcommand{\cF}{\mathcal{F}}
\newcommand{\cG}{\mathcal{G}}
\newcommand{\cH}{\mathcal{H}}
\newcommand{\cL}{\mathcal{L}}
\newcommand{\cM}{\mathcal{M}}
\newcommand{\cMideal}{\cM_\mathrm{ideal}}
\newcommand{\cK}{\mathcal{K}}

\newcommand{\cR}{\mathcal{R}}
\newcommand{\cS}{\mathcal{S}}
\newcommand{\cT}{\mathcal{T}}
\newcommand{\cU}{\mathcal{U}}
\newcommand{\cV}{\mathcal{V}}

\newcommand{\sG}{\mathscr{G}}
\newcommand{\sGClif}{\sG_\mathrm{Clif}}
\newcommand{\Eu}{\cE_{\upu}}
\newcommand{\Gu}{\cG_{\upu}}
\newcommand{\Lu}{\cL_{\upu}}
\newcommand{\Mu}{\cM_{\upu}}
\newcommand{\Ru}{\cR_{\upu}}
\newcommand{\idu}{\id_{\upu}}

\newcommand{\Mideal}{M_\mathrm{ideal}}
\newcommand{\LmdL}{\Lambda^{\mathrm{L}}}
\newcommand{\LmdR}{\Lambda^{\mathrm{R}}}

\newcommand{\Gt}{\widetilde{\cG}}
\newcommand{\Gtu}{\widetilde{\cG}_{\upu}}
\newcommand{\Ghu}{\overline{\cG}_{\upu}}
\newcommand{\Bh}{\overline{\cB}}
\newcommand{\mhat}{\widehat{\mathbf{m}}}
\newcommand{\bsigma}{\boldsymbol{\sigma}}

\newcommand{\Du}{D}

\begin{document}

\title{Comparing randomized benchmarking figure with average infidelity of quantum gate-set}
\date{\today}
\author{Jiaan Qi}
\affiliation{Department of Physics, National University of Singapore}
\email{jiaanq@u.nus.edu}
\author{Hui Khoon Ng}
\affiliation{Yale-NUS College, Singapore}
\affiliation{Centre for Quantum Technologies, National University of Singapore}
\affiliation{MajuLab, CNRS-UCA-SU-NUS-NTU International Joint Research Unit, Singapore}
\email{huikhoon.ng@yale-nus.edu.sg}
\homepage{http://quantum-nghk.commons.yale-nus.edu.sg}

\begin{abstract}
Randomized benchmarking (RB) is a popular procedure used to gauge the performance of a set of gates useful for quantum information processing (QIP). Recently, Proctor et al.~[\href{https://doi.org/10.1103/PhysRevLett.119.130502}{Phys.~Rev.~Lett. \textbf{119}, 130502 (2017)}] demonstrated a practically relevant example where the RB measurements give a number $r$ very different from the actual average gate-set infidelity $\epsilon$, despite past theoretical assurances that the two should be equal. Here, we derive formulas for $\epsilon$, and for $r$ from the RB protocol, in a manner permitting easy comparison of the two. We show in general that, indeed, $r\neq \epsilon$, i.e., RB does not measure average infidelity, and, in fact, neither one bounds the other. We give several examples, all plausible in real experiments, to illustrate the differences in $\epsilon$ and $r$. Many recent papers on experimental implementations of QIP have claimed the ability to perform high-fidelity gates because they demonstrated small $r$ values using RB. Our analysis shows that such a statement from RB alone has to be interpreted with caution.
\end{abstract}
\maketitle

\section{Introduction}
Randomized benchmarking (RB)~\!\cite{Emerson2005,Emerson2007,Silva2008,Knill2008,Dankert2009,Magesan2011,Magesan2012,Moussa2012,Wallman2014,Barends2014a,Cross2015,Carignan-Dugas2015,Granade2015,Helsen2017,Zhang2017,Combes2017,Wallman2017} 
is a group-symmetrization procedure designed to measure properties of quantum gates. In its most widely used form, our focus here, RB is intended for estimating the average infidelity of gates in the $n$-qubit Clifford group, a useful figure-of-merit for quantum information processing (QIP).
Compared with quantum process tomography \cite{Chuang1997,Poyatos1997}, RB is resource-efficient, and robust against state preparation and measurement errors. Simple to implement, RB has been used in a  large variety of experiments, from benchmarking gate performances \cite{Knill2008,Brown2011,Gaebler2012,Harty2014,Veldhorst2014,Kawakami2016a,Olmschenk2010,Xia2015a,Ryan2009a,Moussa2012,Park2016a,Rong2015,Laucht2015,Muhonen2015,Chow2009,Chow2010,Corcoles2013,Barends2014,Fogarty2015,Sheldon2016,Willsch2017,Magesan2012a} to creative uses like partial tomography and quantum control \cite{Kimmel2014,Gambetta2012,Wallman2015,Wallman2015a,Ball2016,Kelly2014a}. 

RB involves applying to an input state, a sequence of gates randomly chosen from the Clifford group, subject to the constraint that they compose to the identity operation in the absence of noise. The fidelity between the input and output states is estimated for varying sequence lengths, yielding a single number---the \emph{RB decay rate} $p$---characterizing the exponential decline in fidelity. Past analyses \cite{Magesan2011,Magesan2012} showed that the average gate-set infidelity $\epsilon$ is well-approximated by the \emph{RB number} $r\equiv\frac{d-1}{d}(1-p)$ ($d$ is the dimension of the system). This is confirmed by numerical simulation \cite{Epstein2014} for a variety of noise models.

In the usual (Markovian) description, a noisy gate can be written as $\Gt=\LmdL_\cG\cG$, where $\cG$ is the ideal gate, and $\LmdL_\cG$ is the gate-dependent ``left'' noise.  Standard RB analyses are based on perturbation about a gate-independent $\LmdL$, yielding $r\simeq\epsilon$ if $\LmdL_\cG$ for every gate in the gate-set is close to $\LmdL$, the situation of weakly gate-dependent noise. Recently, however, Proctor et al.~\cite{Proctor2017} demonstrated an example where the numerically observed RB decay rate is far from that predicted by theory. Their example had noise with gate dependence stronger than accommodated by the perturbative approach past analyses. As their example was practically relevant, an analysis tolerant to stronger gate-dependent noise was needed. Ref.~\cite{Proctor2017} gave an improved analysis for $r$; separately, Wallman \cite{Wallman2017} provided a formula for $r$ accurate to high order in the gate-dependence of the noise, using perturbative analysis about a different form for $\Gt$. Both works gave theoretical predictions for $r$ that fit the observed RB behavior well. 

However, it is now unclear what $r$ measures, as $\epsilon\neq r$ remains for the example of \cite{Proctor2017}. Proctor et al.~argued that $\epsilon$ is an unmeasurable quantity as it is gauge-dependent, and discussed a gauge-independent alternative. Yet, the average infidelity precisely quantifies how close the actual situation is to the experimenters' ideal description. It is a key figure of merit used in many theoretical and experimental papers on QIP. The question then remains: How is $\epsilon$ related to $r$?

Here, we first provide a much more transparent derivation of Wallman's formula for the RB decay rate, making a simplifying but, in practice, insignificant assumption. We then derive an exact expression for the average gate-set infidelity, one that allows for easy comparison with the RB number. This demonstrates very clearly that $\epsilon\neq r$ in general, i.e., that RB does not measure average infidelity. In fact, one does not even provide a bound for the other. We illustrate our results with naturally motivated examples. Recent QIP experiments have reported, using RB, impressive fidelity of their gate-sets (see, for example, \cite{Brown2011,Harty2014,Barends2014,Rong2015}). In light of our work, those conclusions should be re-examined.

\section{Math preliminaries}
Let $\cB(\cH)$ denote the set of operators on $\cH$, the $d$-dimensional Hilbert space of a quantum system. Every $A\in\cB(\cH)$ is a vector $\kket{A}$ in a vector space $\cV$ with inner product $\bbkk{A}{B}\equiv\tr(A^\dagger B)$. Vectors in $\cV$ are represented by $d^2$-element column vectors by choosing an orthonormal (ON) basis $\{O_\alpha\}_{\alpha=0}^{\Du}$ ($\Du\equiv d^2-1$) for $\cB(\cH)$ with $\bbkk{O_\alpha}{O_\beta}=\delta_{\alpha\beta}$. $\bbkk{O_\alpha}{A}$ gives the $\alpha$th entry of the column representing $\kket{A}$. An ON basis with $O_0=\id/\sqrt{d}$ is called a $\id$-basis; a Hermitian basis is a $\id$-basis with $O_\alpha=O_\alpha^\dagger~ \forall \alpha$.

Superoperators---linear maps taking operators to operators---belong to $\cB(\cV)$.
A choice of ON basis $\{O_\alpha\}$ for $\cB(\cH)$ permits representing $\cE$ as a $d^2\times d^2$ matrix with entries $\cE_{\alpha\beta}\equiv \bbkk{O_\alpha}{\cE(O_\beta)}$.  
A trace-preserving (TP) superoperator is represented using a $\id$-basis as
\begin{equation}\label{eq:EMatrix}
\cE\,\widehat{=}\,{\left(
\begin{array}{c|ccc}
1 & 0\\ \hline 
\vb t & {\color{white}{\rule{0.1cm}{0.5cm}}}\phantom{M}\Eu\phantom{M}\\[0.5ex]
\end{array}\right)},
\end{equation}
where $\vb t$ is a $\Du$-element column vector, $0$ is the $\Du$-element zero row vector, and $\Eu$ is, a $\Du\times \Du$ matrix, is the ``unital part" of $\cE$. 
A quantum gate $\cG$ is a unitary superoperator acting as $\cG(\,\cdot\,)=G(\,\cdot\,)G^\dagger$ with unitary $G\in\cB(\cH)$. A quantum channel (e.g., $\cG$ and $\LmdL_\cG$) is a completely positive (CP) TP superoperator.

Every superoperator can be associated with a $d^4$-dimensional vector via a vectorization map. 
The vectorization can be defined in a basis-independent manner (see Appendix A), but for concreteness, we use a Hermitian basis for operators. Superoperator $\cE\in\cB(\cV)$ is then represented by a matrix, and we associate $\cE$ with the vector $\vc(\cE)$ formed by stacking the columns of that matrix. Let $|\cE)\equiv \vc(\cE)$ and $(\cE|\equiv [\vc(\cE)]^\dagger$. Note the identity
\begin{equation}\label{eq:vecid}
|\cE\cF\cG)=(\cG^\transpose\otimes \cE)|\cF),~ \textrm{for any }\cE,\cF,\cG\in\cB(\cV),
\end{equation}
where $\cG$ and $\cE$ on the right-hand side are the matrices representing the superoperators, and $\transpose$ denotes the transpose. For CP $\cE$, its matrix is real, as is $|\cE)$. 
We need two specific superoperators,
\begin{equation}\label{eq:BMatrices}
\cB_{0} \widehat{=}\,{\left(
\begin{array}{c|ccc}
1 & 0\\ \hline
&\\[-1em]
0 & \phantom{M}0\phantom{M}\\[0.5ex]
\end{array}\right)}
,\enskip 
\cB_{1}\widehat{=}\,{\left(
\begin{array}{c|ccc}
0 & 0\\ \hline 
0 & {\color{white}{\rule{0.1cm}{0.5cm}}}\frac{1}{\sqrt D}\idu\\[0.5ex]
\end{array}\right)},
\end{equation}
where $\idu$ is the $\Du\times \Du$ identity. Let $|0)\equiv \vc(\cB_{0})$ and $|1)\equiv \vc(\cB_{1})$. 
Often, we vectorize not the entire $\cE$, but its unital part $\Eu$, yielding a $D^2$-dimensional vector $|\Eu)=\vc(\Eu)$. We also use $|\Eu)$ to denote  a vector in the $d^4$-dimensional space of superoperators, with the additional components set to zero.
Note that $|\Eu)=(\id\otimes \Eu)|\idu)=\sqrt D(\id\otimes \Eu)|1)$, for any $D\times D$ $\Eu$.

\section{Average fidelity}
The imperfect implementation of an ideal gate $\cG$ is $\Gt$, a quantum channel. 
For a pure input state $\psi\equiv\ket\psi\bra\psi$, the \emph{gate fidelity} is $F_\cG(\psi) \equiv F{\bigl(\cG(\psi),\Gt(\psi)\bigr)}$, where $F(\psi,\rho)\equiv \tr(\psi\rho)\in[0,1]$ is the usual (squared) fidelity. 
The \emph{average gate fidelity}, $F_\cG$, is $F_\cG(\psi)$ averaged over $\psi$, under the Fubini-Study measure $\dd\psi$ \cite{Bowdrey2002,Nielsen2002}, or, equivalently, the Haar measure for a fiducial state $\psi_0$:
\begin{align}
F_\cG &\equiv \int\!\!\dd\psi F_\cG(\psi)
\!= \!\!\int\!\! \dd{\psi}\, \bbra{\psi} \LmdR_\cG \kket{\psi}\nonumber\\
&= \bbra{\psi_{0}}\! \!\int\! \!\dd\cU\, \cU^{\dagger}\! \LmdR_\cG \cU \kket{\psi_{0}},
\end{align}
where $\LmdR_\cG\equiv \cG^\dagger\Gt$ with $\cG^\dagger(\cdot)\equiv G^\dagger (\cdot) G$, the adjoint of $\cG$. $\LmdR_\cG$ is the ``right" noise, i.e., $\Gt=\cG\LmdR_\cG$.

In QIP, one is concerned with the performance of a set of gates $\sG$. The quantity of interest becomes the \emph{average gate-set fidelity}, defined as a double average over $\psi$ and gates in $\sG$:
\begin{equation}
F_\sG \equiv \avg{F_\cG}=\bbra{\psi_{0}} {\left(\int\! \dd\cU \, \cU^\dagger \! \LmdR_\sG  \, \cU\right)} \kket{\psi_{0}},
\end{equation}
where $ \LmdR_\sG\equiv \avg{\LmdR_\cG}$, with $\avg{\,\cdot\,}\equiv\frac{1}{\vert\sG\vert}\sum_{\cG\in\sG} (\,\cdot\,)$, is the gate-set average right noise. 

The expression within the parentheses is the \emph{twirl} of $\LmdR_\sG$ with respect to the unitary group \cite{Bennett1996}. The twirl of $\cE$ gives the superoperator
\begin{equation}
\cT(\cE)\equiv\int\dd \cU\, \cU^\dagger \cE U\,\widehat{=}\,{\left(
\begin{array}{c|ccc}
1 & 0\\ \hline 
0 & \phantom{M}q_\cE\idu\phantom{M}\\[0.5ex]
\end{array}\right)},
\end{equation}
where $q_\cE=\Tr(\Eu)/\Du$. For a CPTP $\cE$, $q\in[-\frac{1}{\Du}, 1]$, 
with $q=1$ when $\cE$ is the identity.
We thus have $F_\cG=\left[(d-1)q_\cG+1\right]/d$, where $q_\cG=\Tr{(\cG^\dagger\Gt)_\upu}/\Du$, and the average gate-set fidelity for $\sG$ is $F_\sG=[(d-1)q_\sG+1]/d$, with
\begin{equation}\label{eq:qG}
q_\sG\equiv\avg{q_\cG}=\frac{1}{\Du}\Tr{\avg{(\cG^\dagger\Gt)_\upu}}=\frac{1}{\Du}\Tr\avg{\Gu^\dagger \Gtu}.
\end{equation}
The average gate-set \emph{infidelity} is $\epsilon\equiv 1-\cF_\sG=$ \mbox{$[(d-1)/d](1-q_\sG)$}. Note that $\cT(\cdot)=\avg{\cG\,\cdot\,\cG^\dagger}$ for $\cG\in\sGClif$, the $n$-qubit Clifford group \cite{Dankert2009}, hinting at a link between $F_\sG$ and RB of Clifford gates.

\section{RB decay rate}
A standard RB sequence comprises $m+1$ gates,
\begin{equation}
\cS_{m+1}\equiv \Gt_{m+1} \Gt_m \cdots \Gt_2 \Gt_1\equiv \Gt_{m+1:1}.
\end{equation}
$\cG_i$ are uniform-randomly chosen from $\sGClif$ such that $\cG_{m+1:1}=\id$; $\Gt_i$s are the imperfect $\cG_i$s applied in reality. $\cS_{m+1}$ acts on an input state $\psi_0$ and the fidelity of the output with $\psi_0$ is measured. This is repeated for many randomly chosen $\cS_{m+1}$s to estimate $F_{m+1}\equiv \bbra{\psi_0}\avgS{\cS_{m+1} } \kket{\psi_0}$ ($\avgS{\cdot}$ denotes the average over sequences) for many $m$ values. $F_{m+1}$ is fitted to a mtodel $f(m)=ap^m+b$. The obtained value of $p$ is the RB decay rate; $a$ and $b$ are $m$-independent constants. The form of $f(m)$ was derived \cite{Magesan2011, Magesan2012} for gate-independent left noise $\LmdL_\cG=\LmdL$ same for all $\cG\in\sGClif$,. Weak gate-dependence in the noise was incorporated perturbatively by considering $\LmdL_\cG=\LmdL+\delta\cG$ with $\delta\cG$ small.

As noted in the introduction, the assumption of weakly gate-dependent left noise may not hold well enough in practice for a perturbation about $\LmdL$ to work. Instead,
Wallman \cite{Wallman2017} computed $p$ for noise with weak gate-dependence in the following sense: $\Gt=\cL\cG\cR+\delta\cG$, where $\cL$ and $\cR$ are $\cG$-independent, and $\delta\cG$ is small. This amounts to a perturbation about $\cL(\cdot)\cR$, rather than $\LmdL$. Wallman's formula gives $p$ accurate to $m$th-order in $\norm{\delta\cG}$, giving greater tolerance to gate dependences in the noise. Here, we give a simple derivation of Wallman's result under the assumption $\Gt_1=\cG_1$, i.e., the first gate in $\cS_{m+1}$ is noiseless. This is a mild assumption for the usual case of large $m$. In return, we gain an intuitive understanding of the formula for $p$ (see also a different derivation in the recent article \cite{Merkel2018}). 

We begin with $\avgS{\cS_{m+1}}$. The constraint $\cG_{m+1:1}=\id$ permits free choice of the last $m$ gates, and the first gate is fixed: $\cG_1=\Gt_1=\cG_2^\dagger \cG_3^\dagger \ldots\cG_{m+1}^\dagger$. 
The independent choices of $\cG_2,\ldots,\cG_{m+1}$ entail independent group averages, so
\begin{equation}
\avgS{\cS_{m+1}} = \avg{\Gt_{m+1} \avg{\Gt_m \cdots \avg{\Gt_2 \cG^{\dagger}_2} \cdots \cG^{\dagger}_{m}} \cG^{\dagger}_{m+1}}.
\end{equation}
Vectorizing both sides, and repeatedly applying Eq.~\eqref{eq:vecid}, we have
\begin{align}
&|\,{\avgS{\cS_{m+1}}}) \\
=&\,\avg{\cG_{m+1}\!\otimes \Gt_{m+1}}\vc(\avg{\Gt_m \cdots \avg{\Gt_2 \cG^{\dagger}_2} \cdots \cG^{\dagger}_{m}})\nonumber\\
=&\, \avg{\cG_{m+1}\!\otimes \Gt_{m+1}} \cdots \!\avg{\cG_2\otimes \Gt_2} \, |\id)=\cM^m |\id),\nonumber
\end{align}
with $\cM\equiv \avg{\cG \otimes \Gt}$, noting that $\cG^\dagger =\cG^\mathrm{T}$.
Now, $F_{m+1}=(\psi_0|\avgS{\cS_{m+1}})=(\psi_0|\cM^m|\id)$, where $|\psi_0)\equiv\vc(\kket{\psi_0}\bbra{\psi_0})$. 
The behavior of $F_{m+1}$ is determined by the eigenstructure of $\cM$.

One can show (see Appendix B) that $\cMideal\equiv \avg{\cG \otimes \cG}=|0)(0|+|1)(1|$. 
$\cM$ is close to $\cMideal$ for weak noise.
For unital noise [$\Gt(\id)=\id$], $\cM=|0)(0|+\Mu$, where $\Mu\equiv\avg{\Gu\otimes\Gtu}$ involves only the unital parts. For weak noise, $M\equiv \Mu$ is not defective, and we write its eigendecomposition as
\begin{equation}\label{equ:decomposition}
M=\sum_{i=1}^D p_i\frac{|L_i)(R_i^\dagger|}{(R_i^\dagger|L_i)},
\end{equation}
where $|L_i)$ and $|R_i^\dagger)$ are normalized right and left eigenvectors of $M$, respectively, with $(R_i^\dagger|L_j)=\delta_{ij}(R_i^\dagger|L_i)$, and $p_i$ are the corresponding eigenvalues, with \mbox{$p_1\geq p_2\geq \ldots$}. Consequently,
\begin{equation}
F_{m+1}=\sum_ip_i^m\frac{(\psi_0|L_i)(R_i^\dagger|\id)}{(R_i^\dagger|L_i)}+b,
\end{equation}
where $b\equiv (\psi_0|0)(0|\id)$, an $m$-independent constant. For weak noise, $M\simeq \Mideal\equiv(\cMideal)_\upu= |1)(1|$, so one expects $p_1\simeq 1$, while $p_{i\geq 2}\simeq 0$. Then, $F_{m+1}\simeq a p_1^m+b$, where $a\equiv (\psi_1|L_1)(R_1^\dagger|\id)/(R_1^\dagger|L_1)$. This is of the expected form $f(m)=ap^m+b$, identifying the RB decay rate $p$ as $p_1$, the largest eigenvalue of $M$. The same conclusion holds for nonunital noise; see Appendix C.

Keeping only $p_1$ amounts to the approximation 
\begin{align}
M&\simeq p_1\frac{|L_1)(R_1^\dagger|}{(R_1^\dagger |L_1)}\\
&=\frac{D p_1}{(\!R_1^\dagger|L_1\!)}(\!\id\otimes L_1\!)\Mideal(\!\id\otimes R_1\!)\equiv M^{(0)}. \nonumber
\end{align}
$M^{(0)}$ can be viewed as the unital part of $\cM^{(0)}=\avg{\cG\otimes\Gt^{(0)}}$ with $\Gt^{(0)}=\cL\cG\cR$, where $\cL$ and $\cR$ are, up to a multiplicative factor, unital maps with $L_1$ and $R_1$ as their respective unital parts. Then, $M^{(0)}=\frac{1}{D}|\Lu)(\Ru^\dagger|$ and
\begin{equation}
p=p^{(0)}\equiv \Tr(M^{(0)})=\frac{1}{D}(\cR^\dagger|\cL).
\end{equation}
One can regard the above analysis as the zeroth-order case of a perturbative analysis that accounts for weakly gate-dependent noise about the gate-independent noise $\Gt^{(0)}=\cL\cG\cR$, i.e., $\Gt=\cL\cG\cR+\delta \cG$ with $\delta\cG$ small. Wallman's analysis carries out precisely this perturbation. 

\section{$p$ versus $q$}
We want to compare the RB decay rate $p$ with the average fidelity parameter $q\equiv q_{\sGClif}$. For any $M=\avg{\Gu\otimes\Gtu}$,
\begin{align}
\label{eq:p}p&=p_1=\textrm{largest eigenvalue of }M,\\
\textrm{and }~\label{eq:q}q&=\frac{1}{\Du}\Tr(\avg{\Gu^\dagger \Gtu})=\rbra{1}M\rket{1},
\end{align}
using Eq.~(\ref{eq:vecid}) in writing $q$. Eq.~\eqref{eq:q} is exact, valid for any $\Gt$. We derived Eq.~\eqref{eq:p} by disregarding the small eigenvalues; Wallman \cite{Wallman2017} showed that the result is in fact accurate to $m$th order in $\delta \cG$. Clearly, unless the eigenvector of $M$ with the largest eigenvalue is $|1)$, $p\neq q$.

To explore the difference between $p$ and $q$, it suffices to consider $\Gt^{(0)}=\cL\cG\cR$ with gate-independent and unital $\cL$ and $\cR$. We drop all $0$ superscripts and write $\Gt=\cL\cG\cR$ and $M=\frac{1}{D}|\Lu)(\Ru^\dagger |$. In this case,
\begin{align}\label{equ:qp}
p&=\frac{1}{\Du}(\Ru^\dagger|\Lu)\nonumber\\
\textrm{and}\quad q&=\frac{1}{\Du}(\Ru^\dagger |1)(1|\Lu).
\end{align}
Let $\alpha\equiv\norm{\Lu}\norm{\Ru^\dagger}/\Du>0$, where $\Vert\cE\Vert \equiv \sqrt{\Tr(\cE^\dagger \cE)}= \sqrt{(\cE|\cE)}\equiv\Vert|\cE)\Vert$. 
Define the unit vectors $|L)\equiv |\Lu)/\norm{\Lu}$ and $|R^{\dagger})\equiv |\Ru^\dagger)/\norm{\Ru^\dagger}$. $\beta\equiv \rbrk{R^{\dagger}}{L}$ denotes their overlap, and $|R^{\dagger})=|L)+\sqrt{1-\beta^2}|L^\perp)$ where is a unit vector with $(L^\perp|L)=0$. Then, $M=\alpha \rdyad{L}{R^\dagger}$, and 
\begin{align}
\frac{p}{\alpha}=\beta,\qquad \frac{q}{\alpha}&=x_1(\beta x_1+\sqrt{1-\beta^2}x_2),\label{eq:q0}
\end{align}
where $x_1\equiv (L|1)$ and $x_2\equiv (L^\perp|1)$ are real numbers with $x_1^2+x_2^2\leq 1$.
When $x_1=1$, $p$ and $q$ are equal. This has $x_2=0$ and $|L)=|1)$, corresponding to $\cL$ being the identity and nontrivial right noise $\cR=\LmdR$. This is the situation of standard RB analyses \cite{Magesan2012}. Actually, \cite{Magesan2012} used a gate-independent left noise $\LmdL$, but since $\bbra{\psi_0}\cT(\avg{\LmdL_\cG})\kket{\psi_0}=\bbra{\psi_0}\cT(\avg{\LmdR_\cG})\kket{\psi_0}=F_\sG$, they arrived at the same conclusion as for a gate-independent right noise.

\begin{figure}
\includegraphics[width=\columnwidth]{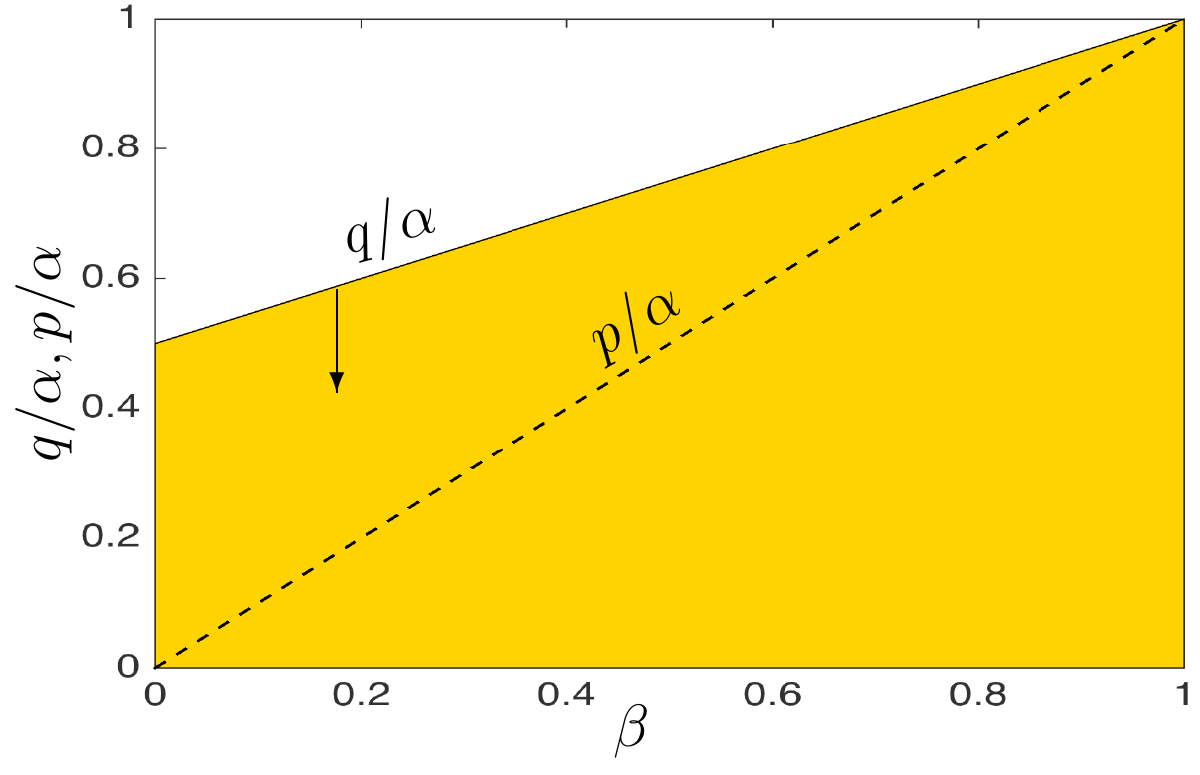}
\caption{\label{fig:1} Plot of $q/\alpha$ and $p/\alpha$ versus $\beta$. The shaded area indicates the possible range of $q$ values for each value of $\beta$; the dotted line marks the value of $p/\alpha$.}
\end{figure}

More generally, $p$ and $q$ are unequal. 
For weak noise, $\cL$ and $\cR$ are close to the identity, and $\beta$ is close to, but less than 1. We hence focus on $\beta\in[0,1]$. For fixed $\alpha$ and $\beta$, $p=\alpha\beta$ is fixed. $q$, however, depends on $x_1$ and $x_2$, determined by how $L$ and $R$ are related to the identity. Eq.~\eqref{eq:q0} gives, for fixed $\alpha$ and $\beta$,
\begin{equation}\label{eq:qlim}
0\leq \frac{q}{\alpha}\leq\frac{1}{2}(1+\beta).
\end{equation}
The limits are attainable: $q/\alpha=0$ when $\Tr(L)=0$, with $R$ chosen to achieve the desired $\beta$; $q/\alpha =(1+\beta)/2$ for $x_1=\cos(\theta/2)$ and $x_2=\sin(\theta/2)$ where $\cos\theta\equiv\beta$.
Fig.~\ref{fig:1} shows the range of values for $q/\alpha$ (shaded area); $p/\alpha$ is single-valued, indicated by the dotted line. Clearly, not only are $p$ and $q$ unequal, $p$ is also neither an upper nor lower bound for $q$. 

Eq.~\eqref{eq:qlim} can be translated into a relation between the average gate-set infidelity $\epsilon$ and the RB number $r$:
\begin{equation}
\epsilon\ge \frac{d-1}{2d}(1-\alpha)+\frac{1}{2}r.
\end{equation}
For the single-qubit case, $\alpha\leq 1$ (see Appendix D), so that
\begin{equation}
\epsilon\ge \frac{1}{2}r.
\end{equation}
From this, we see that $\epsilon$ is large if $r$ is large. However, when $r$ is small, one cannot say anything about $\epsilon$. RB can thus be a witness for poor gate performance, but cannot be used to demonstrate high fidelity.

\section{Examples}
One does not need exotic $\cL$ and $\cR$ to see the variety of behavior in the relative sizes of $p$ and $q$. $\cL$ and $\cR$ as Pauli channels suffice: $\cL(\cdot)=(1-\Sigma l)(\cdot)+l_x X(\cdot)X+l_y Y(\cdot)Y+l_z Z(\cdot) Z$, with $\Sigma l\equiv l_x+l_y+l_z\leq 1$, $l_x,l_y,l_z\geq 0$; $\cR$ is similarly defined, with analogous parameters $s_x,s_y,s_z$, and $\Sigma s$. In this case, one finds $q=p+\frac{4}{9}{\left[(\Sigma l)(\Sigma s)-3\boldsymbol{l}\cdot\boldsymbol{s}\right]}$, where $\boldsymbol{l}\equiv(l_x,l_y,l_z)$ and $\boldsymbol{s}\equiv(s_x,s_y,s_z)$. When $\cL$ and $\cR$ are both depolarizing channels, i.e., $\boldsymbol{l}=(l,l,l)$ and $\boldsymbol{s}=(s,s,s)$, $q=p$. When $\cL$ and $\cR$ are both dephasing channels, with $\boldsymbol{l}=(0,0,l)$, and $\boldsymbol{s}=(0,0,s)$, $q=p-\frac{8}{9}l s\leq p$. When $\cL$ is a dephasing (in $Z$) channel with $\boldsymbol{l}=(0,0,l)$ while $\cR$ dephases in the $X$ and $Y$ directions equally so that $\boldsymbol{s}=(\frac{1}{2} s,\frac{1}{2} s,0)$, we have $q=p+\frac{4}{9}l s\geq p$.

A particularly stark example with $q<p$ is when $\cL={\cR}^\dagger=\cU$ a unitary (see recent independent discussion of this in \cite{Carignan-Dugas2018}).
$q\neq 1$ whenever $\cU$ is different from the identity. However, since $\Vert\Lu\Vert^2=\Vert\Ru\Vert^2=\Du$, Eq.~\eqref{equ:qp} gives $p=1$.
That $p=1$ is clear from the RB sequence for this case: $\cS_{m+1}=\cU\cG_{m+1}\cU^\dagger \cU\cG_m \cU^\dagger \ldots \cU\cG_1 \cU^\dagger =\id$. This is not a pathological case: A basis misalignment gives such a situation.  

By the same argument, two unitarily related noisy gate-sets $\{\Gt\}$ and $\{\Gt'\equiv \cU\Gt\cU^{\dagger}\}$ for some gate-independent $\cU$ have the same $p$ but different $q$ values. If $\Gt=\cL\cG\cR$, then $\Gt'=\cL'\cG\cR'$ with $\cL'=\cU\cL$ and $\cR'=\cR\cU^\dagger$. Since $(\Lu|\Ru^\dagger)=(\Lu'|\Ru'^\dagger)$, $\{\Gt\}$ and $\{\Gt'\}$ have identical $p$ values. However, the primed and unprimed vectors have different relationships with $|1)$, and hence different  $x_1$ and $x_2$, giving different $q$ values. Such unitary freedom can result in drastically different $\epsilon$ and $r$.
Suppose $\Gt= \cG\Lambda_\cG(\theta)$ for a CPTP noise $\Lambda_\cG(\theta)$ that differs from the identity by $\sim\theta\ll 1$. Eq.~\eqref{equ:qp} gives $\epsilon\sim1-q\sim\theta$. Now, if there is a $\cG$-independent $\cU(\theta)$ such that, for every $\cG$, $\cU(\theta)\Gt\cU(\theta)^\dagger=\cG \Lambda_\cG'(\theta^2)$, where $\Lambda_\cG'(\theta^2)$ differs quadratically in $\theta$ from the identity, we have $r\sim1-p \lesssim \theta^2$, drastically different from $\epsilon$.  

Another concrete case where $1-p$ and $1-q$ differ significantly is that of \cite{Proctor2017}. There, they numerically explored RB for noisy single-qubit Clifford gates with coherent noise. Specifically, every gate is composed from an order-1 sequence of noisy ``primitive" gates, namely, $\pi/2$ rotations along the $x$ or $y$ axes, with the same unitary right noise $\cU_P(\theta)=\upe^{-i \frac{1}{2}\theta \widehat{\mathbf{m}} 
\cdot\boldsymbol{\sigma}}$, where $\mhat$ is a unit vector in the Bloch sphere (in \cite{Proctor2017}, $\widehat{\mathbf{m}}=\widehat{\mathbf{e}}_z$), and $\theta\ll 1$. Then, $\Gt_k=\cG_k\cU_k$ where $\cU_k\equiv\upe^{-i \frac{1}{2}\theta_k \widehat{\mathbf{m}}_k 
\cdot\boldsymbol{\sigma}}$ is a gate-dependent unitary for unit vector $\widehat{\mathbf{m}}_k$ and $\theta_k=\theta g_k(\theta)(\ll 1)$ some function of $\theta$. Straightforward computation (see Appendix E) gives $\epsilon\simeq\frac{1}{6}\avg{\theta_k^2}$. For the example of \cite{Proctor2017}, $\avg{\theta_k^2}=\frac{3}{2}\theta^2$, so $\epsilon\sim \theta^2$, as seen numerically in \cite{Proctor2017}. To get $p$, we solve the eigenvalue equation for $M\equiv\Mideal+\sum_{n=1}^\infty\theta^nM^{(n)}$ order-by-order in $\theta$ to obtain $p=1+\sum_{n=0}^\infty \theta^np^{(n)}$. For the example of \cite{Proctor2017}, we find $p^{(1)},p^{(2)},p^{(3)}=0$, and $p^{(4)}=-\frac{233}{864}$ is the first nonzero term, giving $r=\frac{1}{2}(1-p)\sim\theta^4$, as numerically observed in \cite{Proctor2017}. That $r\!\sim\!\theta^4$ while $\epsilon\!\sim\!\theta^2$ seems peculiar, however, to this example. General unitary noise on single-qubit gates, i.e., $\Gt_k=\cG_k\cU_k$, need not show this behavior. For example, when $\theta_k=a_k\theta$ for constants $a_k$, one finds (see Appendix E) $\epsilon=\frac{1}{6}\avg{a_k^2}\theta^2$ and $r\simeq \frac{1}{2}p^{(2)}\theta^2$ with a generally nonzero $p^{(2)}$. Of course, if the noise is such that $p^{(2)}=0$, then we again have $r=O(\theta^3)\ll \epsilon$. Also, if the $\theta$ in the example of \cite{Proctor2017} is random (e.g., Gaussian distributed), $p^{(2)}$ no longer vanishes, and one has $r\sim\theta^2$, just as for $\epsilon$.

The recent article \cite{Carignan-Dugas2018}, interestingly, showed that, for the single-qubit case, there is always a $\cU$ that allows ``correction'' of this kind such that $\epsilon = r $ with error up to $O(r^2)$; for larger systems, Ref.~\cite{Carignan-Dugas2018} gives some numerical evidence for such a possibility, though no proof was provided. If this conclusion is indeed extendible to higher dimensions, one might then argue that, if one can always account for the deviation of $\epsilon$ from $r$ by such a unitary change of basis, after which, $\epsilon$ becomes as small as $r$, then perhaps such a deterioration in the average fidelity---without the $\cU$ fix---is a spurious one.

Whether this is the case depends on one's goals. If the goal is to use RB to deduce the actual size of the average fidelity, defined as the comparison of  the noisy implementation to the ideal description, then, such a fix is of no avail; one simply cannot use RB to measure the average fidelity for a particular implementation. If  the goal is, instead, to have a figure of merit for gates, comparable across different platforms, and modulo these unitary problems, then RB works perfectly. However, one needs to be careful to take the step of having good gate performance as measured by RB to making the statement that one can do high-fidelity computation. The unitary deviations, thought of as basis misalignment between the implementation and the description, have experimentally observable consequences: The interpretation of computational and measurement results relies on our description of the setup, and any such misalignment yields answers that deviate from the expected ones.

\section{Conclusion}
We derive formulas for the average fidelity parameter $q$ and the RB decay rate $p$ in a manner permitting easy comparison of the two. Resulting inequality relations between the average infidelity $\epsilon$ and the RB number $r$ show neither one bounds the other. We give several examples, relevant for real experiments, to illustrate differences between $\epsilon$ and $r$. For the single-qubit case, $\epsilon$ is large whenever $r$ is large, but the converse is not true. This means that RB can tell us when the qubit gates have poor fidelity, but cannot be used to certify high fidelity. The main assumption of past analyses, that the gate-noise is well-approximated by a gate-independent \emph{left} noise, i.e., $\Gt\simeq \LmdL\cG$, is simply one that can fail in practice; the general situation requires both gate-independent left and right noise, i.e., $\Gt\simeq \cL\cG\cR$. Recently, many experimental implementations of QIP gates have claimed the achievement of high-fidelity quantum gates, as evidenced by small $r$ values using RB. Our analysis here indicates that one needs to be careful about drawing such conclusions from $\epsilon$ measured by RB.

\section*{\hspace*{-0.5cm} Acknowledgments}
The authors thank Chai Jing Hao, Berge Englert, Jun Suzuki, and Zheng Yicong for insightful discussions. This work is supported by the Ministry of Education, Singapore (through grant number MOE2016-T2-1-130). HKN is also supported by Yale-NUS College (through a start-up grant). The Centre for Quantum Technologies is a Research Centre of Excellence
funded by the Ministry of Education and the National Research Foundation of Singapore. JQ is supported by a PhD scholarship from the Department of Physics, NUS.

\bigskip
\bigskip
\appendix
\section{Basis-independent vectorization map}
When we discussed vectorization of superoperators, we chose a Hermitian basis for operators and used that to associate with every superoperator, the vector formed by stacking the columns of the matrix representing that superoperator in the chosen basis. This is what is usually known as the Choi-Jamiolkowski isomorphism \cite{Choi1975,Jamiolkowski1972}. The vectorization map can, however, be defined in a basis-independent manner, as we explain here.

We first define the involution $*:\kket{A}\in\cV\mapsto\kket{\overline A}\in\cV$, such that $\forall\kket{A},\kket{B}\in\cV$ and complex scalars $c_a$s, $\bbkk{\overline B}{\overline A}=\bbkk{A}{B}$ (*-symmetry), and $*\Bigl(\sum_ac_a\kket{A}\Bigr)=\sum_ac_a^*\kket{\overline A_a}$ (anti-linearity).
The $*$-map is defined for superoperators by $*(\kket{A}\bbra{B})\equiv \kket{\overline A}\bbra{\overline B}$, and extended to all $\cB(\cV)$ by anti-linearity. We write $*(\cE)\equiv \overline \cE$ for $\cE\in\cB(\cV)$. The vectorization map is then the linear map
\begin{equation}
\vc\bigl(\kket{A}\bbra{B}\bigr)\equiv *\bigl(\kket{B}\bigr) \otimes \kket{A} = \kket{\overline{B}} \otimes \kket{A}.
\end{equation}
One can check that $\vc(\cdot)$ is basis independent because $*$ is basis independent.
Note that $(\overline \cG)^\dagger = \overline{(\cG^\dagger)}$, and the identity Eq.~\eqref{eq:vecid} reads as
\begin{equation}
|\cE\cF\cG)=(\overline \cG^\dagger \otimes \cE)|\cF),\quad \cE,\cF,\cG\in\cB(\cV).
\end{equation}
In this basis-independent form, $\cM$ from the main text becomes $\cM=\avg{\overline{\cG}\otimes\Gt}$.

\section{Formula for $\cMideal$}
We first introduce an ON basis for superoperators: $\{\cB_{\mu\nu}\}_{\mu,\nu=0}^{D}$, with $\cB_{\mu\nu}$ represented as $d^2\times d^2[=(D+1)\times (D+1)]$ matrices using a Hermitian operator basis,
\begin{align}
\cB_{00}&\widehat{=}\,{\left(
\begin{array}{c|ccc}
1 & 0~0\ldots\\ \hline 
0 & \\[-1ex]
0 & 0\\[-1ex]
\vdots&\\
\end{array}\right)},\quad
\cB_{ij}\widehat{=}\,{\left(
\begin{array}{c|ccc}
0& 0~0\ldots\\ \hline 
0 & \\[-1ex]
0 & B_{ij}\\[-1ex]
\vdots&\\
\end{array}\right)},
\nonumber\\
\cB_{01}&\widehat{=}\,{\left(
\begin{array}{c|ccc}
0& 1~0\ldots\\ \hline 
0 & \\[-1ex]
0 & 0\\[-1ex]
\vdots&\\
\end{array}\right)},\quad
\cB_{0D}\widehat{=}\,{\left(
\begin{array}{c|ccc}
0& \ldots 0~1\\ \hline 
0 & \\[-1ex]
0 & 0\\[-1ex]
\vdots&\\
\end{array}\right)},\nonumber\\
\cB_{10}&\widehat{=}\,{\left(
\begin{array}{c|ccc}
0& 0~0\ldots\\ \hline 
1 & \\[-1ex]
0 & 0\\[-1ex]
\vdots&\\
\end{array}\right)},\quad
\cB_{D0}\widehat{=}\,{\left(
\begin{array}{c|ccc}
0& 0~0\ldots \\ \hline 
\vdots &\\[-1ex]
0 & 0\\[-1ex]
1&\\
\end{array}\right)},
\end{align}
where $i,j=1,\ldots, D$, and $\{B_{ij}\}$ forms a Hermitian basis for the unital sector. Note that $\cB_0$ of Eq.~\eqref{eq:BMatrices} is just $\cB_{00}$, and $\cB_1=\cB_{11}$, with vectorized versions $|0)$ and $|1)$, respectively. Observe that $\Tr(\cB_{00})=1$, $\Tr(\cB_{11})=\sqrt D$, and all other $B_{\mu\nu}$s are traceless. 

We also need to state the twirl of an arbitrary superoperator more precisely. The twirl of $\cE$, as defined in the main text, is $\cT(\cE)\equiv\int\dd\cU\,\cU\cE\cU^\dagger$, and gives the following map on operators,
\begin{equation}
[\cT(\cE)](\,\cdot\,)=\frac{1}{d}[t(\cE)-q(\cE)]\tr(\,\cdot\,)\id_\dd+q(\cE)(\,\cdot\,),
\end{equation}
where $t(\cE)\equiv \frac{1}{d}\tr\bigl(\cE(\id_\dd)\bigr)$, and $q(\cE)\equiv\frac{1}{D}\Tr(\Eu)$ as in the main text. Equivalently, one can write
\begin{equation}
\cT(\cE)=t(\cE)\cB_{00}+\sqrt{D}q(\cE)\cB_{11}.
\end{equation}
Note that $t(\cE)=1$ when $\cE$ is TP. 

Now, we can compute the matrix element
\begin{align*}
(\cB_{\mu\nu}|\cMideal|\cB_{\mu'\nu'})&=(\cB_{\mu\nu}|\avg{\overline{\cG}\otimes\cG}|\cB_{\mu'\nu'})\nonumber\\
&=(\cB_{\mu\nu}|\avg{\cG \cB_{\mu'\nu'}\cG^\dagger}),
\end{align*}
for $\cG\in\sGClif$. $\avg{\cG \cB_{\mu'\nu'}\cG^\dagger}=\cT(\cB_{\mu'\nu'})=t(\cB_{\mu'\nu'})\cB_{00}+\sqrt D q(\cB_{\mu'\nu'})\cB_{11}$, with
\begin{align}
t(\cB_{\mu'\nu'})&=\frac{1}{d}\tr\bigl(\cB_{\mu'\nu'}(\id_\dd)\bigr)=\delta_{\mu'0}\delta_{\nu'0}\nonumber\\
q(\cB_{\mu'\nu'})&=\frac{1}{D}\Tr\bigl((\cB_{\mu'\nu'})_\upu\bigr)=\frac{1}{\sqrt D}\delta_{\mu'1}\delta_{\nu'1}.
\end{align}
Putting the pieces together, we have
\begin{align}
&\quad (\cB_{\mu\nu}|\cMideal|\cB_{\mu'\nu'})\\
&=\delta_{\mu'0}\delta_{\nu'0}(\cB_{\mu\nu}|\cB_{00})+\delta_{\mu'1}\delta_{\nu'1}(\cB_{\mu\nu}|\cB_{11})\nonumber\\
&=\delta_{\mu'0}\delta_{\nu'0}\delta_{\mu0}\delta_{\nu0}+\delta_{\mu'1}\delta_{\nu'1}\delta_{\mu1}\delta_{\nu1},\nonumber
\end{align}
so that $\cMideal = |\cB_{00})(\cB_{00}|+|\cB_{11})(\cB_{11}|=|0)(0|+|1)(1|$ as stated in the main text.

\section{$\cM$ and the RB decay rate for nonunital noise}
In the main text, we looked only at the simplest case of unital noise, i.e., $\Gt(\id)=\id$, so that $\cM =|0)(0|+M$. Here, we discuss the general case of nonunital $\Gtu$, and show that we arrive at the same conclusion, that $p$ is the largest eigenvalue of $M=\avg{\Gu\otimes\Gtu}$. 

Under a $\id$-basis, the unitary gate $\cG$ can be written as $\cG=\cB_{00}+ \Gu$  where $\Gu$ is to be thought of, when appropriate, as a matrix of the same size as $\cG$, with vanishing first row and column. As a result, $\cM$ is split into two disjoint blocks,
\begin{equation}
\cM=\left(\Bh_{00}\otimes \avg{\Gt} \right) + \left(  \avg{\Ghu\otimes\Gt} \right)
\end{equation}
In the ideal case, the first and second block become $|0)(0|$ and $|1)(1|$, respectively, giving $\cMideal=|0)(0|+|1)(1|$ as before.

More generally, $\cM=\cMideal+\cK$, with $\cK$ a small perturbation. 
From the Bauer-Fike theorem \cite{Bauer1960}, the shift of each eigenvalue in going from $\cMideal$ to $\cM$ is bounded by the norm of $\cK$, small for weak noise. Hence, $\cM$ has only two eigenvalues close to 1, and all other eigenvalues are close to zero. For the situation of RB, one raises $\cM$ to a large power $m$, so the small eigenvalues quickly die off, leaving only the close-to-1 eigenvalues. 

Now, the eigenvalues $\lambda$ of $\cM$ satisfy the equation $\det(M-\lambda\id)=0$. 
Both $\cG$ and $\Gt$ are TP superoperators. In a $\id$-basis, they can be represented as
\begin{equation}
\cG~\widehat{=}~{\left(
\begin{array}{c|c}
1& 0\\[-0.2em]
\hline
0&\Gu
\end{array}
\right)},\quad\textrm{and}\quad 
\Gt~\widehat{=}~{\left(
\begin{array}{c|c}
1&0\\[-0.2em]
\hline
\rule{0cm}{0.4cm}\mathbf{t}&\Gtu
\end{array}
\right)}.
\end{equation}
Consequently, $\cM$ can be represented as the matrix
\begin{equation}
\cM~\widehat{=}~{\left(
\begin{array}{c|c}
\begin{array}{c|c}
1&0\\[-0.2em]
\hline
\rule{0cm}{0.4cm}\avg{\mathbf{t}}&\avg{\Gtu}
\end{array}&0\\
\hline
0&\rule{0cm}{0.5cm}\avg{\Gu\otimes\Gt}\\[1ex]
\end{array}
\right)}
\end{equation}
From this, keeping in mind that $\Gt$ has its first row as $1,0,\ldots,0$, it is not difficult to see that the determinant of $\cM-\lambda\id$ takes the form 
\begin{align}
\det(\cM-\lambda\id)&=(1-\lambda)\det(\avg{\Gtu}\\
&\quad -\lambda\id)(-\lambda)^D\det(\avg{\Gu\otimes\Gtu}-\lambda\id)\nonumber
\end{align}
Setting this to zero, we find that $\cM$ has an eigenvalue equal to 1, and $D$ zero eigenvalues. 
Since $\avg{\Gtu}$ is small (it vanishes in the ideal case when $\Gtu=\Gu$), the other close-to-1 eigenvalue must come from the $\det(\avg{\Gu\otimes\Gtu}-\lambda\id)$ piece, i.e., it must be the largest eigenvalue $p_1$ of $M=\avg{\Gu\otimes\Gtu}$, just as in the unital case.

Because of the nonzero nonunital part $\mathbf{t}$ of $\Gtu$, the eigenvector corresponding to the eigenvalue 1 of $\cM$ will not simply be $|0)$ as we had for the unital case. Nevertheless, we can still write
\begin{equation}
\cM=|\widetilde 0)(\widetilde 0| + p_1 \frac{|L_1)(R^\dagger_1|}{(R^\dagger_1|L_1)}+\textrm{(small terms)},
\end{equation}
with $|\widetilde 0)$ the eigenvector of $\cM$ with eigenvalue 1, and $|L_1)$ and  $(R_1^\dagger|$ are the right and left eigenvectors of  (the nonunital) $\cM$ with eigenvalue $p_1$. As before, when computing $F_{m+1} = (\psi_0|\cM^m|\id)$, $|\widetilde 0)$ contributes only to the $m$-independent $b$. We see the dominant decay coming from the $p_1^m$ term, and hence, just as in the unital case, we identify the RB decay rate to be
\begin{equation}
p=p_1=\textrm{largest eigenvalue of }M.
\end{equation}

\section{$\alpha\leq 1$ for the single-qubit situation}
With $\alpha$ as defined in the text, we have $M=\alpha |L)(R^\dagger|$. $\alpha$ is the (nonzero) singular value of $M$, and 
\begin{equation}
\alpha = \Vert M\Vert_2\equiv \max_{|\cE)}\frac{\Vert M|\cE)\Vert}{\Vert|\cE)\Vert}=\max_{\Vert|\cE)\Vert=1}\Vert M|\cE)\Vert.
\end{equation}
Recall from the main text that $\Vert |\cE)\Vert \equiv \sqrt{(\cE|\cE)}=\sqrt{\Tr(\cE^\dagger \cE)}=\Vert\cE\Vert$ is the Euclidean norm for vectors, or the Hilbert-Schmidt norm for superoperators.
Now, $\alpha=\Vert M\Vert_2= \Vert\avg{\Gu\otimes\Gtu}\Vert_2\leq \avg{\Vert\Gu\Vert_2\Vert\Gtu\Vert_2}$. Here, $\Vert\cE\Vert_2 $ is the analogous quantity for superoperators acting on operators: $\Vert\cE\Vert_2=\max_{\Vert\kket{A}\Vert=1}\Vert\cE\kket{A}\Vert=\max_{\Vert A\Vert=1}\Vert\cE(A)\Vert$. Since $\cG$ is unitary, so is $\Gu$ and $\Vert\Gu\Vert_2=1$. Observe that $\Vert\Gtu\Vert_2=\max_{\tr(A)=0,\Vert A\Vert=1}\Vert\Gt(A)\Vert$, restricted to traceless inputs.
From Ref.~\cite{Perez-Garcia2006} (Theorem 3.2), any CPTP qubit map $\cE$ has 
\begin{equation}
\max_{\tr(A)=0, \Vert A\Vert=1}\Vert \cE(A)\Vert\leq 1.
\end{equation}
Using this, we have $\Vert\Gtu\Vert_2\leq 1$, and thus $\alpha \leq 1$ for the single-qubit case.

\section{Unitary noise for one qubit}
Consider the situation where the elements in the qubit Clifford gate-set have gate-dependent unitary noise, i.e., $\Gt_k=\cG_k\cU_k$, for $\cG_k\in\sGClif$ ($k=1,2,\ldots,24$), and $\cU_k$ is unitary: $\cU_k(\cdot)=U_k(\cdot)U_k^\dagger$ with $U_k\equiv \exp({-\upi\frac{\theta_k}{2}\mhat_k\cdot\bsigma})$, $\mhat_k$ a three-dimensional spatial unit vector. Straightforward algebra gives the unital part of $\cU_k$, compactly written in a three-dimensional dyadic notation as
\begin{equation}
(\cU_k)_\upu=\cos\theta_k\idDyad - \sin\theta_k\mhat_k^{\times}+(1-\cos\theta_k)\mhat_k\mhat_k,
\end{equation}
where $\idDyad$ is the identity dyadic (matrix), $\mhat_k^\times\equiv\mhat_k\times\idDyad$ denotes the antisymmetric \emph{matrix} with matrix elements $(\mhat_k^\times)_{ij}=\epsilon_{ij\ell}(\mhat_k)_\ell$ where $\epsilon_{ij\ell}$ is the completely antisymmetric tensor.
From $(\cU_k)_\upu$, $q$ is easily computed,
\begin{align}
q&=\tfrac{1}{3}\Tr\avg{(\cG_k)_\upu^\dagger (\widetilde\cG_k)_\upu^\dagger}=\tfrac{1}{3}\Tr\avg{(\cU_k)_\upu}\nonumber\\&=\tfrac{1}{3}(1+2\avg{\cos\theta_k}).
\end{align}
For $\theta_k$ small, the typical case of weak noise, $q\simeq 1-\frac{1}{3}\avg{\theta_k^2}$. 

It is useful to think of $\theta_k$ as a function of a small parameter $\theta\ll1$ such that $\theta_k=\theta g_k(\theta)$, where $g_k$ is some function of the small angle $\theta$. For example, one could have $\theta_k=a_k\theta$, where $g_k(\theta)=a_k$ is an order-1 constant. In this case, $q\simeq1-\frac{1}{3}\avg{a_k^2}\theta^2$. In the example of \cite{Proctor2017}, one can take $\theta$ to be the rotation angle for the $z$-axis noise for the primitive $X$ and $Y$ gates. There, $g_k(\theta)$ is generally a complicated function of $\theta$, but with the help of Mathematica, we find $\avg{\theta_k^2}=\frac{3}{2}\theta^2$, so that $q\simeq 1-\frac{1}{2}\theta^2$. Below, we refer to the situation where $\theta_k=a_k\theta$ as case 1, and the example of \cite{Proctor2017} as case 2.

To get $p$, we organize the eigenvalue calculation in powers of $\theta$. We write $M=\Mu=\avg{\Gu\otimes\Gtu}$, $\cU_k$, $p$, $|L)$ and $(R^\dagger|$ in powers of $\theta$:
\begin{align}
\cU&=\sum_{n=0}^\infty\theta^n\cU^{(n)},\quad p=\sum_{n=0}^\infty\theta^np^{(n)},\nonumber\\
M&=\sum_{n=0}^\infty\theta^n M^{(n)}=\sum_{n=0}^\infty\avg{\cG_k\otimes\cG_k\cU_k^{(n)}},\nonumber\\
|L)&=\sum_{n=0}^\infty\theta^n|L^{(n)}),~ (R^\dagger|=\sum_{n=0}^\infty\theta^n({R^\dagger}^{(n)}|.
\end{align}
Note that $\cU^{(0)}=\id$, $M^{(0)}=\Mideal$, $p^{(0)}=1$, $|L^{(0)})=|1)$ and $({R^\dagger}^{(0)}|=(1|$.

We solve the eigenvalue equations $M|L)=p|L)$ and $(R^\dagger|M=p(R^\dagger|$ order by order in $\theta$. For the linear-in-$\theta$ terms, we have
\begin{equation}
M^{(0)}|L^{(1)})+M^{(1)}|L^{(0)})=p^{(0)}|L^{(1)})+p^{(1)}|L^{(0)}),\label{eq:theta1}
\end{equation}
and a corresponding equation for $({R^\dagger}^{(n)}|$. Multiplying Eq.~\eqref{eq:theta1} on the left by $({R^\dagger}^{(0)}|=(1|$ gives
\begin{align}
p^{(1)}&=\frac{({R^\dagger}^{(0)}|M^{(1)}|L^{(0)})}{({R^\dagger}^{(0)}|L^{(0)})}=(1|M^{(1)}|1)\nonumber\\
&=\frac{1}{3}\Tr\avg{\cU_k^{(1)}}.
\end{align}
In case 1, $\cU_k^{(1)}=-a_k\mhat_k^\times$ which is traceless, so $p^{(1)}=0$; for case 2, using Mathematica, one can show that $p^{(1)}$ also vanishes. For both cases then, $|L^{(1)})=M^{(1)}|1)$.

A similar calculation for the $\theta^2$ terms in the eigenvalue equations, taking $p^{(1)}=0$ (true for both cases), gives
\begin{equation}
p^{(2)}=(1|(M^{(1)})^2|1)+(1|M^{(2)}|1).
\end{equation}
In case 1, $\cU^{(2)}=\frac{1}{2}a_k^2(\idDyad-\mhat_k\mhat_k)$ with trace $\Tr(\cU^{(2)})=-\frac{1}{2}a_k^2(3-\mhat_k\cdot\mhat_k)=-a_k^2$, so that $(1|M^{(2)}|1)=\frac{1}{3}\Tr\avg{\cU^{(2)}}=-\frac{1}{3}\avg{a_k^2}$. The other term, $(1|(M^{(1)})^2|1)=\frac{1}{3}\Tr(\avg{\cU_k^{(1)}}\avg{\cG_\ell\cU_\ell^{(1)}\cG_\ell^\dagger})$, is generally unequal to $\frac{1}{3}\avg{a_k^2}$, so one expects $p^{(2)}$ to be nonzero for case 1. Then, $p= 1-(\textrm{constant})\theta^2$, of similar dependence on $\theta$ as $q$. Of course, one can have a noise such that $p^{(2)}$ vanishes, resulting in an $r$ that is of higher-order in $\theta$ than $\epsilon$.
In case 2, for example, one finds $(1|(M^{(1)})^2|1)=\frac{1}{2}$ and $(1|M^{(2)}|1)=-\frac{1}{2}$, so $p^{(2)}=0$.

Continuing on with only case 2, with $p^{(1)},p^{(2)}=0$, the $\theta^3$ term gives
\begin{align}
p^{(3)}&=(1|(M^{(1)})^3|1)+(1|M^{(3)}|1)\\
&\qquad +(1|M^{(1)}M^{(2)}|1)+(1|M^{(2)}M^{(1)}|1).\nonumber
\end{align}
With the help of Mathematica, one has $(1|(M^{(1)})^3|1)=0$, $(1|M^{(1)}M^{(2)}|1)=-\frac{1}{36}$, $(1|M^{(2)}M^{(1)}|1)=-\frac{1}{72}$, and $(1|M^{(3)}|1)=\frac{1}{24}$, the sum of which gives $p^{(3)}=0$. One can continue this process to obtain an expression for $p^{(4)}$ (the first non-zero correction), but one already sees that $p=1+O(\theta^4)$, of a very different $\theta$ dependence than $q$. Indeed, \cite{Proctor2017} observed numerically that $1-p\sim \theta^4$ while $1-q\sim \theta^2$.

In fact, for case 2 where $p^{(1)},p^{(2)},p^{(3)}=0$, $p^{(4)}$ is given by the intuitive formula,
\begin{align}
p^{(4)}&=(1|(M^{(1)})^4|1)+(1|(M^{(2)})^2|1)+(1|M^{(4)}|1)\nonumber\\
&\quad +(1|{\left[M^{(1)}M^{(3)}+M^{(3)}M^{(1)}\right]}|1)\nonumber\\
&\quad +(1|{\left[\!(M^{(1)})^2M^{(2)}\!+\!M^{(2)}(M^{(1)})^2\!\right.}\nonumber\\
&\hspace{3cm}{\left.+M^{(1)}M^{(2)}M^{(1)}\!\right]}|1).
\end{align}
Using Mathematica, one finds $(1|(M^{(1)})^4|1)=\frac{1}{4}$, $(1|(M^{(2)})^2|1)=\frac{205}{864}$, $(1|M^{(4)}|1)=\frac{7}{144}$, $(1|{\left[M^{(1)}M^{(3)}+M^{(3)}M^{(1)}\right]}|1)=-\frac{41}{72}$, and $(1|{\left[(M^{(1)})^2M^{(2)}+M^{(2)}(M^{(1)})^2\right]}|1)=-\frac{17}{72}$, giving altogether, $p^{(4)}=-\frac{233}{864}\neq 0$.

\bibliographystyle{plainnat}
\bibliographystyle{customstyle}
\bibliography{RB-quantum}

\end{document}